\documentstyle[12pt,aaspp4]{article} 

\newcommand{\bq}{\begin{equation}}
\newcommand{\eq}{\end{equation}}

\begin{document} 
\def\refitem{\par\parskip 0pt\noindent\hangindent 20pt}

\title{The First Type Ia Supernovae: An Empirical Approach
to Taming Evolutionary Effects In Dark Energy Surveys from SNe Ia at \emph{z} \boldmath{$>2$}}

\author{Adam G.\ Riess\altaffilmark{1,2} and Mario Livio\altaffilmark{1}}

\altaffiltext{1}{Space Telescope Science Institute, 3700 San Martin
Drive, Baltimore, MD 21218.}
\altaffiltext{2}{Johns Hopkins University, 3400 N. Chalres Street, Baltimore, MD, 21218}

\begin{abstract}

Future measurements of the nature of dark energy using Type Ia
supernovae will require a precise characterization of systematic sources of error. Evolutionary effects remain the most uncertain contributor to the overall systematic error budget. Present plans to probe evolution with cosmology-independent explosion parameters could yield absence of evidence for evolution without definitive evidence of its absence.  Here we show that observations of Type Ia supernovae in the redshift interval
$1.5 < z < 3.0$, where dark energy-dependent effects are relatively negligible,
should provide direct evidence to discern evolutionary effects.  As examples of our approach to constraining evolution, we examine the impact of changing progenitor 
metallicity and age on the degree of potential luminosity evolution.
We show that the observations we
propose can be carried out by existing space telescopes, or ones that are
already under development.

\end{abstract} 
 
\section{Introduction}

The discovery of the apparent acceleration of the Universe from observations of
high-redshift type~Ia supernovae (SNe~Ia; Riess et~al.\ 1998; Perlmutter et~al.\ 
1999) indicates the presence of a repulsive-like gravity on current horizon-size scales. Dubbed 
``Dark Energy,'' but little understood, 
elucidating its nature provides perhaps the biggest challenge in physical 
cosmology. While many powerful observational tools have been focusing
on this problem, including the cosmic microwave background, baryon acoustic 
oscillations, gravitational lensing, the integrated Sachs-Wolfe Effect, and cluster 
counting, observations of SNe~Ia remain one of the premier tools for 
characterizing the nature of dark energy. Measurement for measurement or source for source, SNe Ia provide the greatest precision for constraining dark energy within the epoch 
of its apparent dominance.

Substantial progress in measuring dark energy parameters with SNe~Ia
requires sample sizes of at least a few thousand, well-sampling the
redshift range $0 < z < 2$. Indeed, some new SN~Ia surveys are
already underway with independent collections of $\sim$100 objects
(ESSENCE, CFHT, SDSS, Higher-$z$) and next-generation SN surveys
are currently under design (Pan-starrs, LSST, JDEM). Yet, to achieve the
precision available from the statistics of such sample sizes will
require controlling the systematic sources of error to approximately 
$\sigma\sim1$--2\%  in luminosity, for each subsample of
a $\sim$ 100 SNe~Ia with uncorrelated errors.

Of all suspected sources of systematic uncertainty specific to SNe Ia, 
evolution is perhaps the most difficult to quantify.
The perceived risk of SNe Ia evolving with redshift 
is driven by two factors: (i)~observations of evolution among {\it other} astronomical
objects pressed into service for measuring cosmic expansion (e.g., Sandage and Hardy 1973), and (ii)~unpredictability 
resulting from the complexity of supernova explosions.

Yet SNe~Ia have been shown to be far less susceptible to evolution than
other astronomical objects, and the mechanism by which they are produced provides a sound reason for them to be ``robust.''  The Chandrasekhar limit can be understood from first 
principles and it homogenizes the explosion mass of SNe~Ia to within a few 
percent. Observationally, no other known astronomical object visible at 
cosmological distances is nearly so uniform nor has it a theoretical reason to be.

Still, some differences do occur during the explosions altering their yields of 
radioactive $^{56}$Ni and varying their apparent peak output by $\sigma\sim20$\% to 25\% 
(e.g., Branch and Tammann 1992; Filippenko 1997; Livio 2001; Leibundgut 2001).  
A single, visible parameter related to the speed of the light curve decline (Phillips 
1993; Riess, Press, \& Kirshner 1995, 1996; Tripp 1998) or spectral feature 
strengths (Nugent et~al.\ 1995) is sufficient to reduce the remaining luminosity 
dispersion to $\sim$15\% (7\% in distance).  To date, the residual peak luminosity 
(in excess from the fiducial value) has no apparent dependence on 
properties of the supernova environment (including stellar population age, and
global metallicity), indicating that samples of supernovae are still reliable when 
averaged to a fraction of their dispersion at moderate redshifts (e.g., 
$z<2$; Riess et~al.\ 1998, 1999, Sullivan et~al.\  2003).  However, as the sample 
sizes surpass 100 SNe~Ia at significant look-back time, the tolerance to a 
systematic error caused by evolution shrinks to $\sim{15\% \over \sqrt{100}}\sim$1.5\% and must be 
questioned and demonstrated.

Empirically, the present task of constraining SN~Ia evolution can be
seen as a quest for a second (third, fourth, etc.) observable parameter
with apparent dependence on luminosity.  A \textit{purely} empirical (i.e.,
uninformed) approach to taming SN~Ia evolution is unlikely to be
successful on simple statistical grounds.  Given a set of $>$7 potential 
observables evenly (and crudely) quantified as having observed values 
which are "low", "medium" and "high", a set of a few thousand SNe~Ia
would be insufficient to even sample (or match similar SNe Ia across redshift)
all possible combinations of these parameters (which would exceed a
couple of thousands).

An approach that is tractable, in principle, is to use a quasi-empirical or 
educated-empirical approach, by employing astronomical knowledge and theoretical 
input to investigate the ``likely suspects'' to constrain SN~Ia evolution.
Unfortunately, complete and highly predictive simulations of SNe~Ia have
proved elusive and are currently insufficient to identify all the important
parameters (many of which are not even varied in the simulations, e.g., 
magnetic fields and rotation; see Discussion).  Current modeling suffers
from known gaps in our ability to characterize the physics of flame propagation 
and explosion conditions, insufficient resolution in position and velocity space, 
and an incomplete itemization of energy levels in the calculation of supernova 
photospheres.  These deficiencies are unlikely to be fully remedied in the near 
future (e.g., Niemeyer, Reinecke, Hillebrandt, 2002)

The major problem with the empirical approach of regressing observed SN
parameters with luminosity is that the absence of evidence for evolution
does not provide evidence of absence.

Here we present an alternative, empirical approach, which requires much less 
guidance from theoretical modeling of supernovae. Our suggested approach 
is to {\it increase\/} possible evolutionary effects while simultaneously 
{\it decreasing\/} the dependence of the apparent brightness on uncertain parameters of 
cosmology. 
At $z>2$, the Universe is observed and expected to be dark matter 
dominated and thus the measured SN~Ia distances are relatively insensitive 
to the nature of dark energy (relative to lower redshifts). However, the 
increased look-back time and the concomitant changes to the global (mean) 
chemical abundances and stellar ages are substantial (as we will show), amplifying any 
potential SN~Ia  evolutionary effect above what may be readily detected at 
lower redshifts. Thus, even without a detailed initial knowledge of the cause 
of evolution, measuring a significant sample of SNe~Ia at $z>2$ should 
reveal the potential effects of evolution (e.g., by direct evidence of its 
absence), \textit{independent} of the behavior of dark energy. In \S2 we 
describe the basic measurement. In \S3 we describe the feasibility of this 
measurement with existing or planned observatories, a discussion follows 
in \S4, and our summary and conclusion are presented in \S5.

\section{SNe Ia at \emph{z} \boldmath{$>$} 2, a Route to Amplifying the Evolutionary Effects}

The change in the luminosity distance with redshift 
 is sensitive to the properties of the 
dark energy within the interval $0<z<1.5$ 
but decreasingly so (with increasing redshift), because the ratio of the
dark matter to dark energy density increases rapidly as $(1+z)^{-3w(z)}$.  Thus, whatever 
the value of the equation of state ($w=P/\rho c^2$) of the dark energy,
 the high-redshift Universe must be dark matter dominated.  As a result, increases
in the luminosity distance with redshift become simpler to predict at higher redshifts.

To illustrate this phenomenon, we show in Figure 1 
{\it differences} in the increase of luminosity distance with
redshift for different dark energy parameters between $0<z<1.5$ and
again between $1.5<z<3.0$. The predicted distances for a cosmological constant-type dark 
energy ($w_0,\ w')=(-1.0,0.0)$ and models with dark energy parameters 
in close proximity ($-$1.0,$-$0.3), ($-$1.0,+0.3), ($-$1.2,0.0), ($-$0.8,0.0) 
differ by $\pm$10\% in flux in the range $0<z<1.5$ [here $w_0$ is $w(z=0)$ and $w' \equiv {dw \over dz}|_{z=0} $]. 
However, in the next redshift range, $1.5<z<3.0$, the {\it additional}
distance for these same dark energy models differs by only $\pm$1\% in flux.  
Thus, while the increase in the luminosity
distance in the range $0 < z < 1.5$ depends on \textit{both} 
(possible) evolution and the unknown dark energy model,
 the luminosity distance increase in the range $1.5<z<3.0$ 
is sensitive \textit{only} to (possible) evolution. 
This relative insensitivity to dark energy at higher 
redshifts provides a powerful way to identify possible SN~Ia evolution.

For example, one of the primary global parameters whose cosmic evolution
might affect SN~Ia luminosities is metallicity (see \S4). To illustrate the utility 
of the approach we advocate, we can reconstruct a representative history 
for the evolution of metallicity based on quasar absorption by damped 
Lyman-alpha systems.  The relevant value of the cosmic metallicity is its 
value at the time of the SN~Ia progenitor formation, not the value at explosion. 
In Figure~2 we show the mean cosmic metallicity derived from 
measurements by Kulkarni et~al.\ 2005 and an assumed delay time for 
SNe~Ia of 2--3 Gyr (Strolger et~al.\ 2005; Dahlen et~al.\ 2005).  As shown, 
the mean global metallicity at the time of a SN Ia progenitor formation 
is expected to decrease by 2~decades in the redshift range of 
$0<z<1.5$ resulting in uncertain, and hard to predict, evolution in SN Ia luminosity. 
This is the redshift range where the observed fluxes of 
SNe~Ia are increasingly being used to characterize dark energy 
(Riess et al. 2004, Aldering et al. 2004). 
In the redshift interval, $1.5<z<3.0$, the mean formation 
metallicity inferred from quasars decreases by an {\it additional} 
2~decades (or more for delay times of $\geq 3 $Gyr).  However, 
the change in flux due to increased distance is easily 
predicted from the value of $\Omega_M$ and depends negligibly on
the dark energy model, providing a measurement of the change due to (possible)
luminosity evolution resulting from decreased metallicity.
(An uncertainty in $\Omega_M of \sim 0.01$,
forecast by Frieman et al. 2003 by the era of the 
Planck Surveyor, propagates only a 2\% uncertainty
in flux 
at the middle of this redshift range, thus retaining the required sensitivity 
to evolution.)

Another example of the utility of our approach derives from the necessary evolution (i.e., decrease) in the maximum age of stars and potential SN Ia progenitors with increasing redshift.
White 
dwarfs at higher redshifts necessarily originate from shorter-lived 
and therefore more massive stars. These stars produce CO 
white dwarfs with a higher initial mass and a smaller ratio of carbon-to-oxygen.
(e.g., Becker \& Iben 1980, Dominguez,  H\"oflich \& Straneiro 2001). Some theoretical models (e.g., 
by Dominguez and H\"oflich 2001) predict a change in the peak luminosity 
of roughly 3\% per 1~solar mass change in the progenitor star mass 
due to the decreased energy yield from the decreased $^{56}$Ni mass 
synthesized in the explosion [generally, a higher CO mass is accompanied by a lower C/O ratio (although not monotonically at the low-mass end), and consequently, a lower energy per gram; see also \S4].
As shown 
in Figure~3, in the dark energy-sensitive redshift range ($0<z<1.5$), 
stars with masses as small as 1~to 1.5~solar masses have sufficient 
time to evolve to white dwarfs and may produce the SNe Ia we see. In the dark matter-dominated range 
($1.5<z<3.0$), the minimum mass rises to 2~to 6~solar masses 
depending on the assumed delay (3 to 0.01~Gyr).  According to
 the models of Dominguez et al. (2001),
the evolution in the luminosity across the range $1.5<z<3.0$ is as large (or larger for delay times of $\geq$ 2 Gyr) as for the range $0.0<z<1.5$, for SNe Ia resulting from the 
white dwarfs with the smallest initial mass which can explode.  Thus,
SNe~Ia measurements at dark energy insensitive 
redshifts should provide a valuable, empirical constraint on the degree 
of such evolution.

The empirical approach that we propose has the additional advantage 
that, in principle, it is not even necessary to know the source of possible 
evolution in order to diagnose its influence at $z>1.5$. While the 
look-back time over $1.5 < z < 3.0$ is modest 
($\sim$2~Gyr) compared to the interval over $0 < z < 1.5$ ($\sim$9~Gyr), the 
known change in the cosmic environmental parameters (likely to effect the formation of 
SN Ia progenitors) is as large or larger. 
Consequently, for simple, monotonic evolution, the potential changes in 
luminosity distances that are due to evolution in the range $1.5< z<3.0$ 
provide an \textit{upper limit} to the changes due to evolution that are 
likely to have occurred across the dark energy-sensitive redshift region.

\section {From Principle to Practice}

While the proposed empirical approach to taming possible SN~Ia 
evolution is powerful in principle, here we consider its practicality.

The most conservative approach to evolution would be to assume SNe Ia
are evolving with redshift and to utilize the measurements at $1.5<z<3.0$
to correct for luminosity evolution.  In that case the precision of the measurement
of the evolution would need to match the precision of the dark energy measurement
sought at $z \sim 1.5$.  

  The leverage on evolution at high-redshift will depend on the assembly time of SN Ia progenitors.  Longer developing progenitor systems originate
  from a younger Universe and thus would provide greater
leverage on evolution.  From Figures 1 and 2 we surmise that assembly times of $>$ 2 Gyr
(favored by Strolger et al. 2005 and Dahlen et al. 2005) 
will provide more evolution per SN Ia 
per unit redshift at $1.5<z<3.0$ than $z \sim 1.5$, reducing
the number of SNe Ia required to constrain evolution by a significant factor.  However, a prompt population of SNe Ia (see Manucci et al. 2004) would likely require an equivalent
sample size of SNe Ia in the two redshift intervals.  Ongoing surveys of SNe Ia
are likely to resolve the assembly time of SN Ia progenitors in the next few years.

Thus, we may assume the number of SNe Ia measured at $1.5<z<3.0$
would need to be between a few hundred and a thousand\footnote{In principle the number
can be as small as the sample size of each redshift bin, defined so that between each pair, systematic errors are uncorrelated.  Current descriptions of future, large number SN surveys (e.g., Aldering et al 2004) generally consider errors due to systematics to be uncorrelated between SNe separated by $\Delta z=0.1$ (e.g., the SNAP proposal, Aldering et al. 2004).  The reasoning
for such de-correlation is rather approximate but originates from the change
in bandpasses and instrumentation used with changing redshift (Linder and Huterer 2003, Aldering 2004).  Within one of these bins, the number of SNe Ia which can be
averaged before reaching an assumed systematic error floor of $\sim$ 1\% in flux is 100.  However, the exact number depends critically on the assumed size of the zeropoint error and its correlation
with redshift, neither of which has been well-established.  Errors due to evolution may be correlated across the entire redshift space of the survey thus raising the number of SNe Ia required
to constrain evolution at $z > 1.5$ to our conservative description.}

An alternative, less demanding approach, would be to utilize the SNe Ia at $z>1.5$
as a coarse check on evolution (``sanity check'').  This approach would be sufficient to
discriminate between the possibilities that SNe Ia are not evolving or if evolving, are
doing so well beyond the unrelated precision of dark energy measurements.  For this approach a few dozen SNe Ia would suffice, but would be unable to rule out an unfortunate coincidence, such that the degree
of evolution of SNe Ia happens to match the desired precision of dark energy measurements (and that the Universe, in EinsteinÕs words, maybe Òis maliciousÓ). 

 Thus, let us assume that the
observational challenge is to measure $\sim$100 SNe~Ia (more or less depending on the aforementioned approaches) at 1.5 to 2.5
microns where they will peak at $24.0<K<24.6$ and $24.0<H<25.5$
and decline by $\sim$1~mag at later phases.  The lower end 
of this redshift range is within the reach of \textit{HST} with the IR channel 
of WFC3, for which the limiting factor is thermal background (which limits 
observations to $<1.72$ micron and thus useful SN Ia measurements to $z< 2.2$). 
 The entire range (and higher) is easily within the 
reach of \textit{JWST}, as well as a possible $\sim$6-m class 
\textit{TPF-C} mission equipped with a general, near-IR astrophysics 
camera.

The survey area required to achieve such a sample of SNe~Ia can be 
determined for an assumed star-formation history (SFH) and SN~Ia 
progenitor formation time function.  If we use the SFH from Giavalisco 
et~al.\ (2005) and a formation delay time of 0 to 3~Gyr, we obtain results 
for WFC3 and \textit{JWST} that are given in Table~1.  

\begin{deluxetable}{ccc} 
\tablewidth{0pt}
\tablecaption{Square degrees required to collect 100 SNe Ia at $1.5 < z < 3.0$
         at typical (1~hour) exposure times}
\tablehead{\colhead{Delay (Gyr)}&\colhead{WFC3$^a$ ($H=25$ Vega)}&\colhead{\textit{JWST} 
($K=27$ Vega $\sim$10 nJy)}}
\startdata
    0 &               0.83    &               0.14\nl
    1 &               0.83     &              0.20\nl
    2 &               1.16      &             0.37\nl
    3 &               2.27       &            0.70\nl
    \enddata 
\tablenotetext{a}{$1.5 < z < 2.2$ is the useful redshift range}
\tablenotetext{a}{Rates normalized to match observed from Strolger et al. 2005}
\end{deluxetable}
  
For the field size of WFC3 (2.3 by 2.1 arcminutes), 745 pointings are required 
to cover an area of one square degree and the maximum useful SN~Ia redshift 
is 2.2.  \textit{JWST}'s NIRcam is planned to have twice the field of view, 
and we can see from Table~1 that it is about three times faster at finding 
these SNe; it will therefore yield about six times as many objects in a typical 
exposure with a maximal redshift surpassing six. Thus, 260 pointings will be 
required for \textit{JWST} for a long delay or 50 pointings for prompt 
SNe~Ia.  This (and more) is readily achievable over the lifetime of this observatory. At 
these redshifts, a field will ``refresh'' and provide new SNe~Ia with a 
timescale of $\sim$2 months (risetime * ($1+z$)).

We conclude that a powerful empirical approach to constraining SN~Ia 
evolution, sufficient for the next generation of SN~Ia-based dark energy 
surveys, will be provided by space telescopes and instrumentation already 
under development.  

\section{Discussion}

Characterizing evolutionary effects in SNe~Ia by theoretical modeling
 is not easy, and quantifying them 
is even harder. There are several reasons for these difficulties: (i)~The 
progenitor systems of SNe~Ia have not been identified unambiguously, 
(ii)~The physics of flame propagation is not fully understood, (iii)~Many 
uncertainties still plague the calculations of the evolution towards explosion,
(iv)~Multi-dimensional simulations of all the processes involved are only 
beginning to emerge. Let us discuss briefly, in turn, the resulting ambiguities.

\subsection{Progenitors}

In spite of considerable progress in the understanding of the progenitor systems, 
and fairly strong theoretical arguments suggesting that SNe~Ia originate from 
single-degenerate systems (in which a white dwarf accretes from a normal 
companion; see Livio 2001 for a review), there is no \textit{observational} 
evidence to show that double-degenerate systems do not produce SNe~Ia. 
Since double-degenerate scenarios differ from single-degenerate ones both in 
the delay time (typically less than 1~Gyr in the double-degenerate scenario, 
compared to $\sim$1.5--3~Gyr in the single-degenerate case), and in the 
environment leading to the explosion, the uncertainty in the identification of the 
progenitor system translates into an uncertain evolutionary effect (especially if 
\textit{both} single-degenerate  and double-degenerate systems can lead to 
SNe~Ia).

\subsection{Physics of the Flame}

Delayed detonation models, in which the burning starts as a subsonic 
deflagration that later turns into a supersonic detonation, have been found to be 
reasonably successful in reproducing both the light curves and spectra of 
SNe~Ia (e.g., H\"oflich, Khokhlov, \& Wheeler 1995; Lentz et~al.\ 2000).  The 
problem is that these models still involve two free parameters: the speed of the 
deflagration, and the density $\rho_{tr}$ at the point of transition from 
deflagration to detonation.  Since the amount of $^{56}$Ni produced (and 
consequently the brightness) depends sensitively on $\rho_{tr}$, the 
lack of a detailed understanding of what determines $\rho_{tr}$ (and of 
whether a transition occurs at all!) is a serious deficiency of the models (see  also \S4.5).

\subsection{The Evolution towards Explosion}

Of the many known uncertainties in the theory of binary star evolution, some are 
particularly crucial for SNe~Ia.  For example, the observed rates of SNe~Ia in 
the low-redshift Universe can be reproduced in the single-degenerate scenario, 
if one assumes an important role for a white dwarf wind during the accretion 
process (this avoids the formation of a common envelope and allows for a 
steady increase in the white dwarf mass; e.g., Hachisu, Kato, \& Nomoto 1999). 
Since optically thick winds are driven by a peak in the opacity, which is due to 
iron lines, the model that invokes an important role for the winds necessarily 
predicts a strong dependence of SNe~Ia rates on metallicity (e.g., Kobayashi 
et~al.\ 1998).  Still, the role of these winds remains, at present, in the realm of 
theoretical conjecture.

\subsection{Multi-Dimensional Effects}

While multi-dimensional SNe~Ia simulations have made remarkable progress 
in the past few years (e.g., R\"opke 2005), treating all the relevant physics in 
three dimensions remains a challenge. In some cases, the newly emerging 
results are in conflict with results from spherically symmetric models (see also 
\S4.5).

\subsection{Possible Evolutionary Effects}

In spite of the uncertainties listed above, the risk of 
evolutionary effects appears to 
be high. For instance, since the high-redshift universe is 
characterized by a lower metallicity, we should consider potential 
metallicity-related effects. Similarly, progenitors in the early universe have 
necessarily to be shorter lived, and therefore of a higher mass. Interestingly, 
detailed calculations of the delayed detonation model by Dominguez et~al.\ 
(2001) and Lentz et~al. (2000) show that for a progenitor with a main 
sequence mass of $M_{MS}=5~M_{\odot}$, a change in the metallicity 
from $Z=0.02$ to $Z=10^{-10}$ has little effect (and is not monotonic) on the 
energetics and on the light curve. Due to changes in the line blending by Fe, 
the decline in the metallicity is expected to be accompanied by a decline in 
$B-V$ by $\sim$0.05$^m$. 

Rather different results were found in the three-dimensional calculations of 
R\"opke et~al.\ (2005).  Specifically, these authors found that changing just 
the $^{22}$Ne mass fraction from 0.5 to 3 times solar resulted in a 20\% 
change in the mass of the $^{56}$Ni produced (and thereby in the SN 
brightness). At present, it is not entirely clear what these variations between 
the spherical and three-dimensional models are due to (see also Timmes et~al.\ 
2003). The metallicity may have a more dramatic effect on SNe~Ia \textit{if} 
indeed the white dwarf wind plays an important role in the evolution towards 
explosion. Specifically, the metallicity-dependent conditions for a strong wind 
(Nomoto et~al.\ 2000) predict a lower bound on the core mass, 
$M_{CO}$ (e.g., $M_{CO}\gtrsim0.95~M_{\odot}$ for 
$Z=0.004$), which translates into an upper bound on the $^{56}$Ni mass. 
Consequently, this model predicts an \textit{absence} of bright SNe~Ia at high 
redshifts (lower metallicity environments). In addition, the strong-wind scenario 
predicts a strong decline in SNe~Ia rates at $z>2.5$.  

Perhaps the largest expected evolutionary effect comes from changes in 
$M_{MS}$. At a fixed value of $Z=10^{-3}$, changing the main sequence mass from 
$1.5~M_{\odot}$ to $6.0~M{\odot}$ results in a change in the average C/O ratio 
in the WD (before explosion) from 0.76 to 0.60 and a concomitant reduction in the 
$^{56}$Ni mass (which is the main factor in determining the explosion strength),  
from $0.59~M_{\odot}$ to $0.52~M_{\odot}$ (Dominguez et~al.\ 2001; Umeda 
et~al.\ 1999).  Consequently, in addition to the change in the peak luminosity, the 
expansion velocities are altered as well, with different chemical layers differing in 
their speeds by up to 1500~km~s$^{-1}$. At $Z=0.02$, increasing 
$M_{MS}$ from $1.5~M_{\odot}$ to $7.0~M_{\odot}$ decreases (in the 
models) the peak luminosity (in both $B$ and $V$) by $\sim$0.15$^m$. Again
we should note that three-dimensional calculations (R\"opke et~al.\ 2005) are at 
variance with the results of spherically-symmetric models when it comes to the 
effect of the progenitor's C/O ratio.  The three-dimensional simulations show only 
a small impact of the C/O ratio on the amount of $^{56}$Ni produced.  This may be due to the fact that the three-dimensional models achieved only partial burning of the white dwarf, and no layered chemical structure.

In the single-degenerate scenario, SNe~Ia at $z\sim3$ could also be
expected to be 
\textit{brighter} due to age effects. The reason is that in these models, the lifetime 
of the system is essentially the lifetime of the companion star, which is determined 
by its mass $M_2$. For a younger system (larger $M_2$), the total mass that can 
be transferred to the white dwarf is higher, requiring a smaller white dwarf mass, 
which translates into a higher $^{56}$Ni mass.

Another potential metallicity-dependent evolutionary effect which has not yet been 
explored in detail is related to neutrino cooling. The density at which the ignition of 
carbon occurs depends sensitively on the cooling rate, which in turn depends on 
the local Urca $^{21}$Ne--$^{21}$F process.  At a lower metallicity environment, 
the abundance of $^{21}$Ne is lower, resulting in less efficient cooling. Ignition 
therefore may occur at a lower central density (lower binding energy of the white 
dwarf). As a result, the light curve evolution may be \textit{faster} in such an 
environment.

  Even if luminosity evolution of SNe Ia does occur in the Universe, its effects
may already be included in, if not already explain, the observed and utilized relation between
light curve shape and luminosity.  An indication that this may be true comes from evidence that the specific light curve shape and luminosity of a SN Ia does depend on the characteristics of its host (e.g., elliptical vs. spiral, e.g. Howell 2005).

\section{Summary and Conclusions}

Future, high-precision measurements of dark energy utilizing thousands of SNe Ia
will require a high degree of control of systematic errors.  The most uncertain and hardest to quantify systematic source of uncertainty impacting SNe Ia is
evolution.  Despite the absence of present evidence of SN Ia evolution, some 
effects that depend on the metallicity and age of the progenitor systems appear 
likely.   Without a complete theoretical description of SNe Ia, better agreement among present efforts to model the explosions, or a massive sample of SNe Ia spanning all combinations of all observable explosion characteristics across redshift, absence of evidence of evolution is unlikely to provide evidence of its absence.  

As an alternative approach to this thorny problem, we have shown that observations of SNe~Ia at $1.5<z<3.0$, where the relative dependence on uncertain dark energy parameters is negligible and the change in progenitor environmental parameters is large, should provide a positive identification of evolution if it is significant.  The observations we propose can be carried out by space telescopes that 
already exist, or are already under development.

\clearpage

\begin{figure}[h]
\vspace*{150mm}
\includegraphics{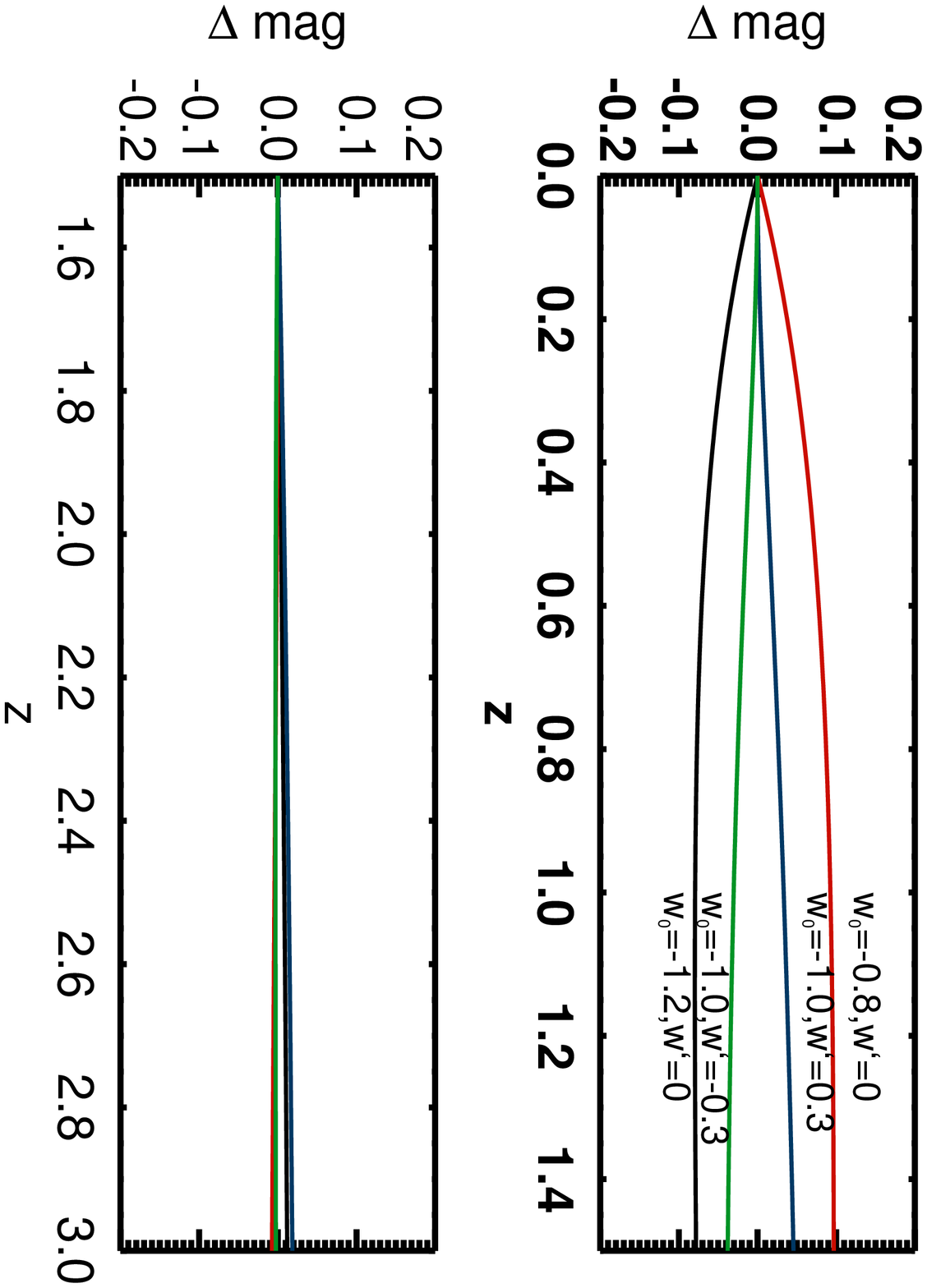}
\caption {Luminosity Distance and Cosmology}
\end{figure}

\begin{figure}[h]
\vspace*{150mm}
\includegraphics{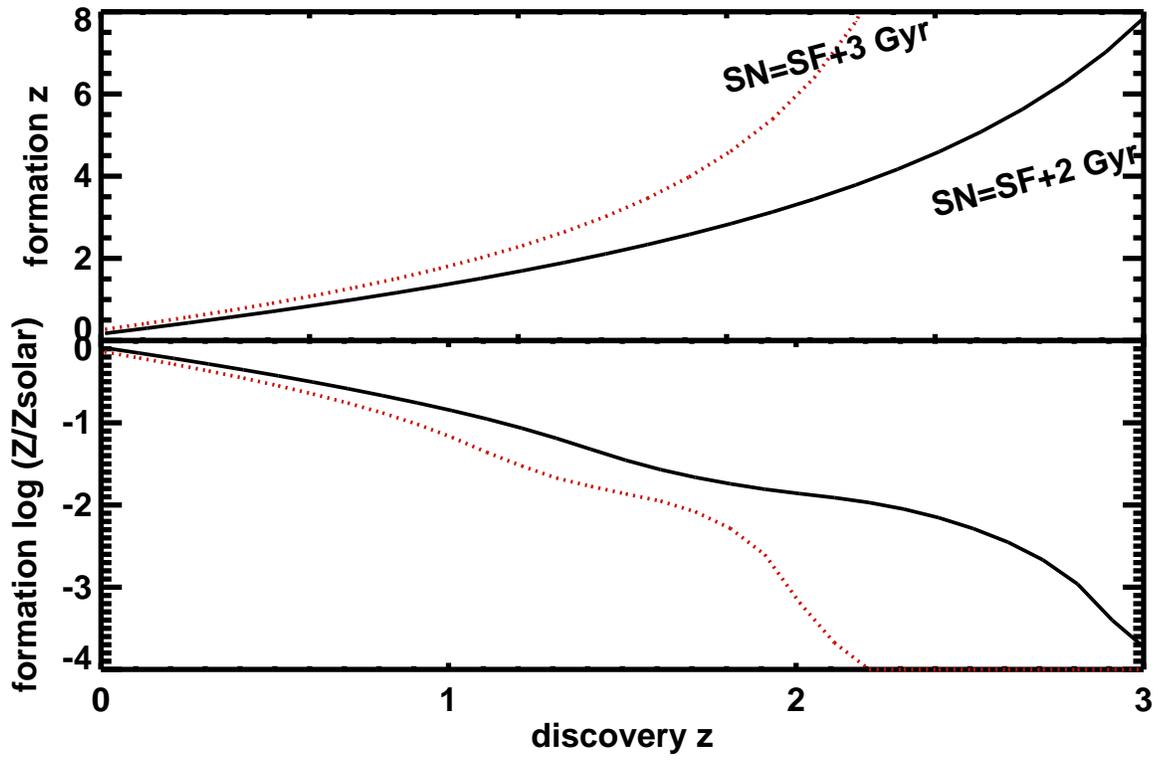}
\caption {Metallicity}
\end{figure}

\begin{figure}[h]
\vspace*{150mm}
\includegraphics{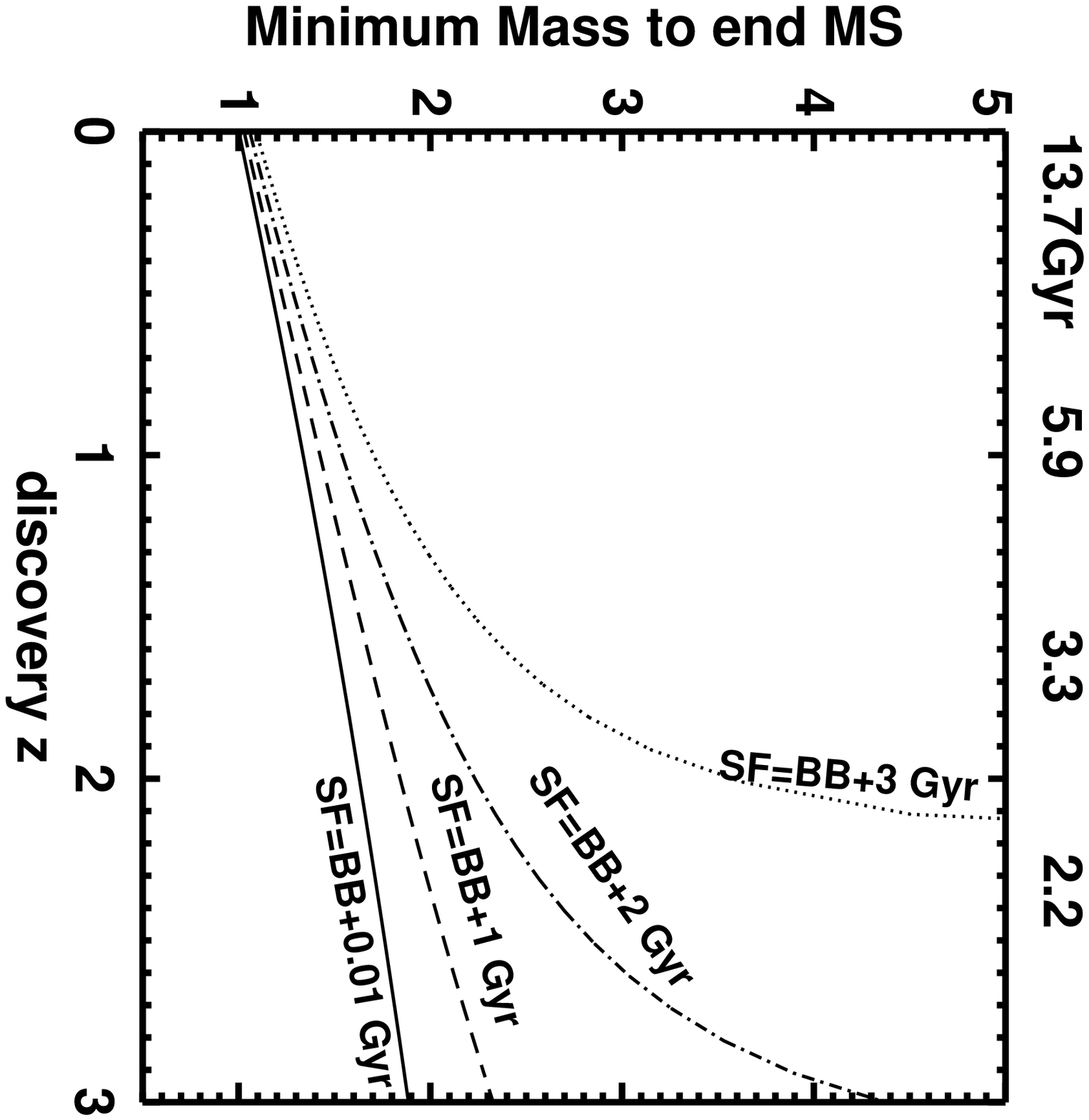}
\caption {Star Age}
\end{figure}

\vfill
\eject

Figure Captions

Figure 1: Dark Energy and Evolution Sensitivity.  At $0 < z < 1.5$, luminosity
distances to SNe Ia are sensitive to differences in the dark energy model, with
a $\pm$ 10\% variation in flux for dark energy parameters currently favored (see upper panel).
However, it is possible that SNe Ia in this range could be altered 
by a significant, unknown  luminosity evolution.  At $1.5 < z <3.0$, variations in the dark
energy model from a fiducial model (cosmological constant) become very small as
the Universe becomes dark matter-dominated (see lower panel).  However, the known cosmic evolution of parameters related to progenitor formation continues in this higher redshift interval
providing the means to diagnose and calibrate the degree of SN Ia evolution affecting
dark energy measurements.

Figure 2: The mean global metallicity of the formation of SN Ia progenitor systems 
as inferred from dampled lyman-alpha systems.  The upper
panel shows the relation between discovery redshift and progenitor formation redshift
depending on the time interval between formation and explosion.  The bottom panel
uses the results from Kulkarni et al. (2005) to show the mean global metallicity
for SNe Ia progenitors as a function of discovery redshift.  As shown,
the variation in global metallicity at $1.5 < z < 3.0$ where SNe Ia distances are relatively insensitive to knowledge of the correct dark energy model, is as larger or larger than at 
$0 < z < 1.5$, where SNe Ia are used to measure dark energy.  These changing dependendies allows
in principle for the breaking of possible degeneracies between SN Ia luminosity evolution
and dark energy.

Figure 3: The minimum main-sequence mass of a progenitor star to evolve to a
white dwarf as a function of the discovery redshift of a SN Ia.  Depending
on the interval between the Big Bang and the formation of the star as indicated, 
the minimum mass star from which the white dwarf originates is a strong function of redshift.
For some SN Ia models (e.g., Dominguez et al. (2001), the initial mass
determines the carbon to oxygen ratio of the white dwarf, a ratio
which can alter the peak luminosity of the SN Ia by 5\% to 15\%.  Thus
expanding the range of redshift where SNe Ia are measured provides
important leveage on this possible evolution scenario. 

\end{document}